\documentclass[final]{aipproc}

\layoutstyle{8x11double}

\usepackage{amsmath}
\usepackage{amssymb}
\usepackage{epsfig}
\usepackage{graphicx}
\usepackage{multirow}
\usepackage{longtable}
\usepackage{lscape}
\usepackage{bbm}
\usepackage{dcolumn}
\usepackage[dvips]{color}
\usepackage{rotating}

\newcommand{\bi}{\begin{itemize}}
\newcommand{\ei}{\end{itemize}}

\newcommand{\be}{\begin{equation}}
\newcommand{\ee}{\end{equation}}
\newcommand{\bea}{\begin{eqnarray}}
\newcommand{\eea}{\end{eqnarray}}

\newcommand{\ldm}{\Delta m_{31}^2}
\newcommand{\sdm}{\Delta m_{21}^2}
\newcommand{\deltacp}{\delta_{\mathrm{CP}}}
\newcommand{\stheta}{\sin^2 2 \theta_{13}}

\newcommand{\ie}{{\it i.e.}}

\newcommand{\eg}{{\it e.g.}}

\newcommand{\fig}{Fig.}

\newcommand{\Ref}{Ref.}
\newcommand{\Refs}{Refs.}

\newcommand{\figu}[1]{\fig~\ref{fig:#1}}

\begin{document}

\title{Performance Comparison: \\ Superbeams, Beta Beams, Neutrino Factory}


\author{Walter Winter}{
  address={Institut f{\"u}r theoretische Physik und Astrophysik, Universit{\"a}t W{\"u}rzburg, D-97074 W{\"u}rzburg, Germany}
}

\begin{abstract}
In this talk, the  performance comparison among superbeams (SB), beta beams (BB), and the Neutrino Factory (NF) is discussed. The ingredients to such a comparison are described, as well as we the optimization and status of BB and NF are addressed. Finally, one example for the performance comparison is shown.
\end{abstract}

\maketitle



The most important channels for the analysis of future long-baseline oscillation experiments are the $\nu_\mu \rightarrow \nu_\mu$ channel to measure the atmospheric parameters, the $\nu_e \rightarrow \nu_\mu$ (BB/NF; ``golden channel''~\cite{DeRujula:1998hd,Cervera:2000kp}) and $\nu_\mu \rightarrow \nu_e$ (SB) channels to measure $\theta_{13}$, the mass hierarchy (MH), and CP violation (CPV). Additional information may, at the NF, be obtained from the ``silver'' $\nu_e \rightarrow \nu_\tau$~\cite{Donini:2002rm,Autiero:2003fu}, ``platinum'' $\nu_\mu \rightarrow \nu_e$~\cite{Bandyopadhyay:2007kx}, and ``discovery'' $\nu_\mu \rightarrow \nu_\tau$ channels~\cite{Donini:2008wz}. Furthermore, neutral current measurements may be useful for new physics searches~\cite{Barger:2004db}. For useful analytical formulas describing the various oscillation channels, see, \eg, \Ref~\cite{Akhmedov:2004ny}. Using the primary $\nu_e \leftrightarrow \nu_\mu$ signal channels for both neutrinos and antineutrinos, often leaves, among multi-parameter correlations~\cite{Huber:2002mx}, three classes of degeneracies which can typically not be resolved: the intrinsic $(\deltacp, \theta_{13})$ degeneracy~\cite{Burguet-Castell:2001ez}, the $\mathrm{sgn}(\ldm)$ degeneracy~\cite{Minakata:2001qm}, and the $\theta_{23}$ (octant) degeneracy~\cite{Fogli:1996pv}, leading to an overall eight-fold degeneracy~\cite{Barger:2001yr}. In order to break this degeneracies, the following methods have been suggested in the literature (see \Ref~\cite{Bandyopadhyay:2007kx} for more details): a)  matter effects, \ie, long baselines, often with multi-GeV energies (\eg, LBNE, T2KK, NF, BB); b) different beam energies or a good energy resolution of the detector (\eg, monochromatic beams, wide-band beams, BB with different isotope pairs); c) the combination of two baselines (\eg, T2KK, NF, BB); d) high statistics (\eg, NF, BB, multi-MW sources, megaton-size detectors); e) the combination of different oscillation channels (\eg, silver and platinum channels at NF, BB plus SB combinations); f) the combination with different experiment classes (\eg, reactor experiments, atmospheric experiments, and astrophysical sources). For example, the ``magic baseline'', a baseline where the dependence on $\deltacp$ disappears independently of energy and the oscillation parameters, turns out to be an extremely efficient degeneracy resolver~\cite{Huber:2003ak}. As another interesting class of experiments, reactor experiments to measure $\theta_{13}$, such as Double Chooz and Daya Bay, will play an important role in the near future. While they are complementary to the long-baseline experiments physics-wise~\cite{Huber:2003pm}, they are also interesting from the conceptual point of view: They are multi-source, multi-detector systems where, because of the used $\bar \nu_e \rightarrow \bar \nu_e$ disappearance channel, systematics is  very important; see, \eg, \Ref~\cite{Mezzetto:2010zi}. Future long-baseline experiments may indeed have similar characteristics: for example, the NF may have four sources (straights), two far detectors, and several near detectors, where the systematics treatment can be similarly important; see, \eg, \Refs~\cite{Giunti:2009en,Tang:2009na}. 

\begin{figure*}[t!]
\includegraphics[width=0.8\textwidth]{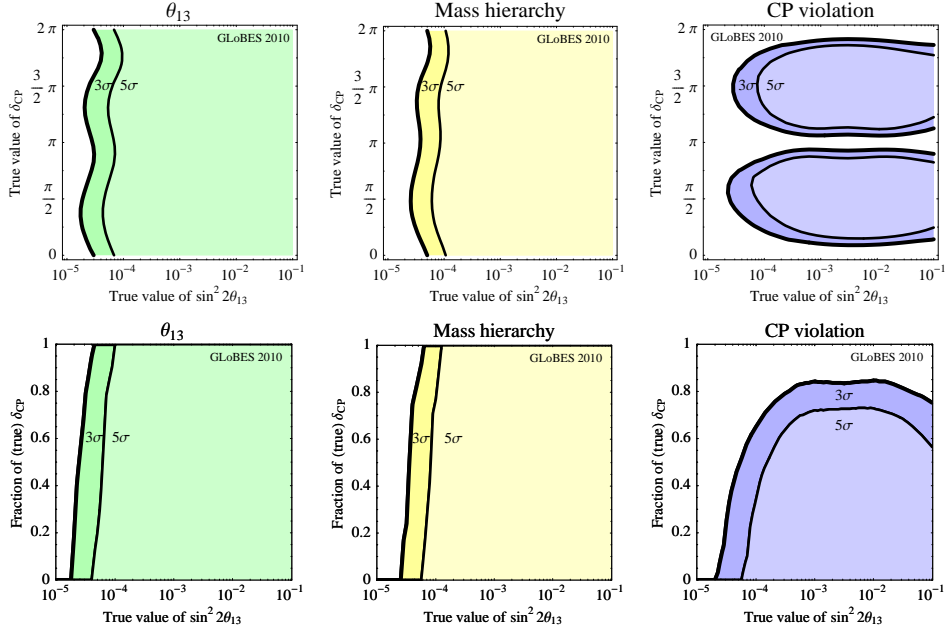}
\caption{\label{fig:heperf}  Performance of the high energy NF (100~kt at 4000~km plus 50~kt at 7500~km) in terms of the $\theta_{13}$, CP violation, and mass hierarchy discovery reaches ($3\sigma$ and $5 \sigma$ CL). The upper row shows the ``raw sensitivity'' as a function of $\stheta$ and $\deltacp$, the lower row the sensitivity as a function of $\stheta$ and fraction of $\deltacp$. See \Ref~\cite{ids} for details.}
\end{figure*}

Apart from standard oscillation physics, future long-baseline experiments will constrain new physics effects. 
 If the new physics originates from heavy mediators which are integrated out, the new physics can be parameterized in the effective operator picture. The lowest possible effective operators affecting the production, propagation, or detection of neutrinos are dimension six operators, suppressed by $v^2/\Lambda^2$ by the new physics scale $\Lambda$ compared to the SM Higgs VeV $v$. At tree level, they can be mediated by heavy neutral fermions, leading to a non-unitary mixing matrix after the re-diagonalization and re-normalization of the kinetic terms of the neutrinos (see, \eg,  \Ref~\cite{Antusch:2009pm} for a short summary), or by scalar or vector bosons, leading to so-called non-standard interactions (NSI; see, \eg, \Ref~\cite{Kopp:2007ne} for the terminology and \Refs~\cite{Antusch:2008tz,Gavela:2008ra} for the theory). At the NF, these effects may be hardly distinguishable because of the leptonic neutrino production by muon decays~\cite{Meloni:2009cg}. Therefore, different neutrino sources will be needed to study these effects.
A very different type of new physics is the oscillation into light sterile neutrinos, because it may lead to oscillation signatures. In this case, the sterile neutrinos are typically light enough to be directly produced by flavor oscillations. Whereas recent MiniBOONE antineutrino results seem to favor a $\Delta m^2 \sim 0.1 - 1 \, \mathrm{eV}^2$~\cite{AguilarArevalo:2010wv}, recent cosmology data may even point to lighter sterile neutrinos~\cite{Hamann:2010bk}. Such very light sterile neutrinos may be tested at future long-baseline experiments if $\Delta m^2 = \mathcal{O}( \ldm)$~\cite{Meloni:2010zr}.

Let us now focus on a purely conceptual comparison among the different experiment classes, including reactor experiments (RE). Then $\theta_{13}$ can, in principle, be measured in RE, SB, BB, and NF. The mass hierarchy can be determined by SB, BB, NF, and, under relatively aggressive assumptions, by RE for large $\theta_{13}$~\cite{Ghoshal:2010wt}. CP violation is measurable in SB, BB, and NF. The atmospheric parameters $\ldm$ and $\theta_{23}$ can be measured by SB and NF independent of $\theta_{13}$, but not at RE and BB. The solar parameters $\sdm$ and, possibly, $\theta_{12}$, may be improved on in a dedicated long-baseline RE. Furthermore, one could study  the requirements for new physics effects from a conceptual perspective. Since some effects, such as some matter NSI, prefer long baselines and high energies  similar to the MSW effect~\cite{Kopp:2008ds}, one may want to cover the MSW resonance energy in the Earth's mantle at about 8~GeV. This would be possible in the NF, and maybe in BB.  Another issue is the separate detection of all neutrino flavors. This requires high neutrino energies to exceed the $\tau$ production threshold. At the NF, for instance, that is in contradiction with $\nu_e$ detection, since the electrons/positrons produce electromagnetic showers for which the charge is difficult to measure for high energies. Therefore, two different muon energies may be required, possibly in a staged approach~\cite{Tang:2009wp}. At the SB and BB, there is not principle problem as long as the beam energies are high enough. 


For the quantitative performance comparison, one has to choose performance indicators. One example are the discovery reaches for $\theta_{13}$, MH, and CPV, illustrated in \figu{heperf} for the NF. For instance, the discovery reach for CPV is defined as the parameter space in (true) $\stheta$ and $\deltacp$, where the CP conserving solutions $\deltacp=0$ and $\pi$ can be excluded at a certain confidence level for any choice of the other oscillation parameters, \ie, the other parameters are marginalized over. The result is depicted in \figu{heperf}, upper right panel. In many cases, for each $\stheta$, the sensitive ranges in $\deltacp$ are then ``stacked'' to the ``fraction of $\deltacp$'' on the vertical axis, see lower right panel. For the $\theta_{13}$ discovery reach, $\theta_{13}=0$ is to be excluded (left panels), and for the MH discovery reach, the other hierarchy is to be excluded (middle panels) at the chosen confidence level for any choice of the other oscillation parameters. As one can see from the figure, the $\theta_{13}$ and MH discovery reaches typically have a simpler structure, because they are mostly limited in the (true) $\stheta$ direction. On the other hand, the CPV discovery will be hard if the true value of $\deltacp$, chosen by Nature, is too close to CP conservation, which leads to two different directions for the optimization (horizontal and vertical in the right panels). Apart from the discovery reaches, deviations from maximal atmospheric neutrino mixing ($\theta_{23}=\pi/4$), see, \eg, \Ref~\cite{Antusch:2004yx}, and the precision measurements of $\theta_{13}$ and $\deltacp$ are interesting performance indicators for future experiments. The precision of $\theta_{13}$ and $\deltacp$ is typically shown as contours (``potatoes'') in the (fit) $\theta_{13}$-$\deltacp$ plane for certain choices of the true values, because these fit contours look like the results which will be published in the future. However, using these, a comparison of experiments difficult. First of all, it is hard to compare two-dimensional fit contours of similar size, but different shape. Second, the shape and size of the contour depends on the true values of $\theta_{13}$ and $\deltacp$. And third, different experiment classes perform differently in different parts of the parameter space, which means that it is not too difficult to choose a parameter set such that a particular experiment looks preferred. Alternative approaches quantify the precision in larger parts of the parameter space, such as the dependence of the $\deltacp$ precision on the true value of $\deltacp$ (``CP patterns''~\cite{Winter:2003ye,Huber:2004gg}). Comparing SB with BB/NF, one can easily see in that approach that SB performs better at (true) $\deltacp=3\pi/2$, the (single baseline) NF between $\pi/4$ and $\pi/2$, \ie, in a different part of the parameter space. For $\theta_{13}$, often the precision of $\theta_{13}$ is shown as a function of the true $\stheta$, where the impact of the true $\deltacp$ is illustrated by bands; see, \eg, \Refs~\cite{Huber:2002mx,Gandhi:2006gu}. For a specific values of $\stheta$, these bands may overlap, which means that one experiment may be better or the other one, depending on the value of $\deltacp$. Because the performance comparison using the precision involves one more degree of freedom, it is not frequently used in the literature, and the performance indicators as in \figu{heperf} are often the standard one.


Let us now discuss the ingredients of the (objective) performance comparison, where one may distinguish:
\begin{enumerate}
\item Factors under machinery control
\item Factors under experiment's control
\item Factors under theorist's control
\item Factors under Nature's control
\end{enumerate}
1. For the comparison machinery, it is important to use similar assumptions for all experiments, such as the 
same oscillation framework, the same (best-fit) oscillation parameters, the same  external input (such as from solar parameters), the same  performance indicators, comparable assumptions on the matter density (profile), the same marginalization techniques, the same $\chi^2$ definition, and a comparable systematics implementation.  This, of course, points towards using the same simulation software, such as the GLoBES software (``General Long Baseline Experiment Simulator'')~\cite{Huber:2004ka,Huber:2007ji}. 2. Here typically information from experimental proposals is taken, such as the beam (source) spectrum or geometry, the detector description (efficiencies, backgrounds, energy resolution), the systematical errors, potentially cross sections, the anticipated luminosity, and the timescale. Especially the latter two quantities should be understood as target values, which implies that existing experiments often deviate significantly from proposed (future) experiments. Thus, in general, the comparison between existing and future experiments is unfair. 3. Unter theorist's control are typically the choice of the performance indicators and parameters, sometimes the systematics implementation (unless specified in detail in the proposal), often the luminosity, which is difficult to compare among the different experiment classes (how to compare protons on target, useful ion decays, useful muon decays?), and design parameters which can be easily changed in the simulation, such as the BB ions, the baseline, and the beam energy. A reasonable strategy for the objective experiment comparison could be only comparing experiments/parameters for which a  cost tag exists, \ie, some experimentalists have thought about feasibility and risk in some detail. However, often new theoretical ideas are tested, for which no proposal or costing exists yet. In this case, the main purpose is the establishment of the physics performance rather than the objective performance comparison. 4. Some factors are under Nature's control. Apart from the true oscillation parameters, for instance, the absolute cross sections can only be predicted in certain models, and, before they are measured, they will therefore have some impact on the absolute performance~\cite{FernandezMartinez:2010dm}, which may be different for different experiment classes.


For the interpretation of the performance comparison, one should also take into account the status and optimization of different experiment classes. The SB are discussed in greater detail in the talk by P.~Huber~\cite{HuberTalk}, which means that we focus on BB and NF here. BB~\cite{Zucchelli:2002sa,Bouchez:2003fy} have been originally proposed with a CERN-based layout. The isotopes $^6$He and $^{18}$Ne were suggested to decay in straights of a storage ring for $\bar\nu_e$ and $\nu_e$ production, respectively. A number of modifications have been proposed later. At the CERN SPS, the isotopes may be accelerated up to $\gamma \sim 100-150$, but higher $\gamma \gg 150$ may be reached in a refurbished SPS or a new accelerator to obtain better sensitivities~\cite{Burguet-Castell:2003vv}. In addition, a new ion production method using a production ring was proposed in \Ref~\cite{Rubbia:2006pi}, which may allow for a highly intense beam from $^6$B and $^{18}$Li decays. The difference between this isotope pair and $^6$He and $^{18}$Ne is the higher endpoint energy, which means that lower boost factors are required to reach the same neutrino energy. Since, however, the total flux is proportional to $\gamma^2$ due to the forward boost, the intensity will be lower for such a beam.  In order to obtain an almost identical beam spectrum, one can show that 
one needs $N_\beta^{(B/Li)} \sim 12 \, N_\beta^{(He/Ne)}$ and $\gamma^{(He/Ne)} \sim 3.5 \, \gamma^{(B/Li)}$,
where $N_\beta$ is the number of useful ion decays; see \Ref~\cite{Agarwalla:2008gf} for details. This means that the lower boost factor has to be payed for by about an order of magnitude more luminous source. From the discussion in \Refs~\cite{Winter:2008cn,Winter:2008dj} it should be clear that a $^6$B/$^{18}$Li-based beam would make sense at CERN for large $\stheta$ if a) a number of useful ion decays/year larger than about $10^{19}$ were reached and b) a sufficiently long baseline for the mass hierarchy determination was chosen (at least about CERN-Gran Sasso).  It is one of the goals of the current Euronu study to investigate the potential of the production ring to reach such high intensities. Apart from the site-specific (SPS-based) discussion, much effort has been spend on alternative BB options, such as with higher $\gamma$'s. For example, in \Ref~\cite{Choubey:2009ks} a $\gamma \gtrsim 350$ BB has been proposed with four ions, altogether at $10^{19}$ useful ion decay per year, send to two detectors, one at 650~km (500~kt water), and one at 7000~km (50~kt iron), which may be a setup similar to the high energy NF.

For the NF~\cite{DeRujula:1998hd,Cervera:2000kp,Geer:1998iz}, there is a currently ongoing  international design study (IDS-NF), which aims for a design report, schedule, cost estimate, and risk until about 2012. Currently, an interim design report  is being prepared,
including details of how the costing will be done. At the current IDS-NF baseline, muons are accelerated to 25~GeV, and the neutrinos are produced by a total of $2.5 \, 10^{20}$ useful muon decays/baseline/polarity in straights of two storage rings, pointed towards two magnetized iron detectors at about 4000~km and 7500~km~\cite{ids}. While the optimization has been so far performed for a green-field setup (see, \eg, \Refs~\cite{Huber:2006wb,Kopp:2008ds}), the review of specific sites is being studied~\cite{HuberPrep}. The long (``magic'') baseline,  which could be a baseline from CERN to India, is an important constituent of this setup. Apart from the degeneracy-free measurement of $\theta_{13}$ and mass hierarchy, it increases the robustness of the experiment with respect to new physics~\cite{Ribeiro:2007ud,Kopp:2008ds}, systematics~\cite{Tang:2009na}, or lower than expected luminosity -- it can be regarded as a risk minimizer. Even the matter density along the baseline could be extracted at the percent level~\cite{Gandhi:2006gu,Minakata:2006am}. An alternative version of the NF with a different detector technology, such as a magnetized TASD (Totally Active Scintillating Detector), and a lower muon energy, is also being studied~\cite{Geer:2007kn,Bross:2007ts,FernandezMartinez:2010zza}. Because of the lower detection threshold of the detector, lower muon energies and shorter baselines are allowed while the performance, especially for large $\stheta$, is maintained or even improved. In practice, especially in the light of the recent optimization of the magnetized iron detector~\cite{Cervera:2010rz,ThesisLaing}, the low and high energy NF may be just two different versions of the same experiment optimized for different detectors and different parts of the parameter space~\cite{HuberPrep}. In fact, if the NF is built in stages, both versions may be found in the final approach~\cite{Tang:2009wp}. 


\begin{figure}[tp]
\includegraphics[width=\columnwidth]{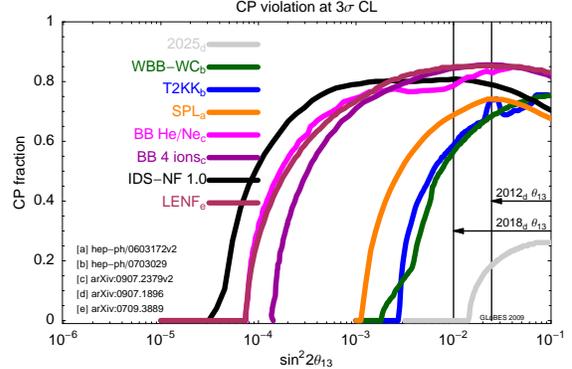}
\caption{\label{fig:euronu2009}
Comparison of the physics reach of different future facilities for the CP violation discovery reach ($3\sigma$).  Curves are taken from: [a] \Ref~\cite{Campagne:2006yx}, [b] \Ref~\cite{Barger:2007jq}, [c] \Ref~\cite{Choubey:2009ks},  [d] \Ref~\cite{Huber:2009cw} and [e]
 \Ref~\cite{Bross:2007ts}. The $\theta_{13}$ sensitivities expected for 2012 and 2018 are shown as vertical lines~\cite{Huber:2009cw}. Figure taken from \Ref~\cite{Bernabeu:2010rz}.  
}
\end{figure}

Finally, in \figu{euronu2009} an example for the comparison of the CPV discovery reach of different experiments is shown. From this figure, one can read off that there are approximately three classes of experiments with three different sets of curves: The existing or planned experiments ``2025'' will most likely not establish CP violation. The SB upgrades ``WBB-WC'', ``T2KK'', and ``SPL'', using megawatt-class targets and megaton-size (water) detectors, can measure CP violation for most of the parameter space if $\stheta \gtrsim 0.01$, which may be established by the next generation of experiments, such as Daya Bay. Note that the SPS-based CERN BB (not shown) roughly falls in that category as well. For very small $\stheta \ll 0.01$, only different versions of the NF or high-$\gamma$ BB ($\gamma \gtrsim 350$) can measure CP violation. All these experiments are based on substantially new technology. Comparing these NF and BB options, in the light of the aspects discussed earlier in this talk, two observations are interesting: First, the low energy NF (LENF), which is the smallest discussed version of the experiment, compares very well with the various high-$\gamma$ BB, which are some of the high-end versions discussed for BB. Second, for the NF, a design study is going on, whereas none of the shown high-$\gamma$ BB are currently experimentally studied. 


In conclusion, the performance comparison among (future) long-baseline neutrino oscillation experiments should be done with care, since there are often very different assumptions hiding behind different curves. That of course does not mean that theorists should not come up with new ideas, for which the physics performance can be instantly compared to existing options using such performance comparisons. In this case, future dedicated studies have to show how these options compare in terms of feasibility, risk, and cost. 

\renewcommand{\refname}{\vspace*{-1.2cm}}


\end{document}